\documentclass[10pt]{article} 

\usepackage[preprint]{tmlr}

\usepackage{amsmath,amsfonts,bm}

\def\eqref#1{equation~\ref{#1}}

\def\1{\bm{1}}

\DeclareMathAlphabet{\mathsfit}{\encodingdefault}{\sfdefault}{m}{sl}
\SetMathAlphabet{\mathsfit}{bold}{\encodingdefault}{\sfdefault}{bx}{n}

\usepackage{hyperref}
\usepackage{url}
\usepackage{subcaption}
\usepackage{graphicx}
\usepackage{float} 

\usepackage{multirow}
\usepackage{array}

\usepackage{algorithm}
\usepackage{algpseudocode}  

\usepackage{hyperref}
\usepackage{url}
\usepackage{booktabs}
\usepackage{wrapfig}   

\title{APEX4: Efficient Pure W4A4 LLM Inference via Intra-SM Compute Rebalancing}

\author{\name Hong Guo \email hong.guo@hpi.de \\
      \addr Hasso Plattner Institute
      \AND
      \name Nianhui Guo \email nianhui.guo@greenbit.ai \\
      \addr GreenBit.AI
      \AND
      \name Weixing Wang \email weixing.wang@hpi.de\\
      \addr Hasso Plattner Institute
      \AND
      \name Jona Otholt  \email jona.otholt@hpi.de\\
      \addr Hasso Plattner Institute 
      \AND
      \name Christoph Meinel \email christoph.meinel@german-uds.de\\
      \addr German University of Digital Science
      \AND
      \name Haojin Yang \email haojin.yanggreenbit.ai\\
      \addr GreenBit.AI}

\def\month{MM}  
\def\year{YYYY} 
\def\openreview{\url{https://openreview.net/forum?id=XXXX}} 

\begin{document}

\maketitle

\begin{abstract}
W4A4 quantization promises full utilization of INT4 Tensor Cores, yet group dequantization overhead on CUDA Cores has driven existing systems to mixed-precision fallbacks. We present the first systematic study of how intra-SM compute balance governs this bottleneck. Through controlled benchmarks across four GPUs from Ampere and Ada architectures, we identify the Tensor Cores to CUDA Cores throughput ratio ($\rho$) as the primary hardware indicator: the W4A4-g128 kernel yields $2.0$--$2.5\times$ speedup on RTX~3090 ($\rho=16$) yet degrades to $0.43$--$0.47\times$ on A100 ($\rho=64$) in compute-bond scenarios, establishing W4A4 viability as platform-dependent rather than universally infeasible. 
Guided by this finding, we build \textbf{APEX4}, which co-designs pure INT4 GEMM kernels with $\rho$-aware granularity adaptation to mitigate the CUDA Cores dequantization bottleneck. 
APEX4 achieves perplexity within 0.63 of FP16 on Llama-2-70B and outperforms W4Ax Atom-g128 by 4.0\%--4.4\% in zero-shot accuracy. Deployed as a drop-in replacement in unmodified vLLM, it delivers up to $1.66\times$ end-to-end speedup on L40S ($\rho=8$), and $1.78\times$ on RTX~3090 ($\rho=16$), $2.09\times$ on A40 ($\rho=16$), while recovering A100 ($\rho=64$) to $1.20$--$1.40\times$ via the mixed-granularity mode.
Our code is available at \url{https://github.com/APEX4-W4A4/APEX4-W4A4}.
\end{abstract}

\section{Introduction}
\label{sec:introduction}

As large language models~(LLMs) continue to scale in parameter count~\citep{llama3,deepseekv3,qwen3}, low-bit quantization \citep{GPTQ,AWQ,QUIP,Quarot,turboquant} has emerged as a critical pathway for efficient LLM inference. Existing schemes, W8A8~\citep{smoothquant}, W4A16~\citep{marlin}, and W4A8~\citep{QQQ,Qserve}, occupy different accuracy-efficiency trade-off points. On the A100, INT4 Tensor Cores (TC) deliver 1248 TOPS versus 312 TFLOPS for FP16~\citep{nvidia_ampere_whitepaper}, a $4\times$ peak throughput advantage that positions W4A4 as an attractive operating point for compute-bound GEMM workloads.
However, the practical realization of this theoretical advantage remains elusive. Many existing W4A4~\citep{ATOM,COMET} systems still rely, to varying degrees, on mixed-precision computation in their kernel designs. QServe~\citep{Qserve} concludes that W4A4 cannot deliver speedup on Ampere and retreats to W4A8.

\begin{wrapfigure}{r}{0.5\textwidth}
    \centering
    \includegraphics[width=0.5\textwidth]{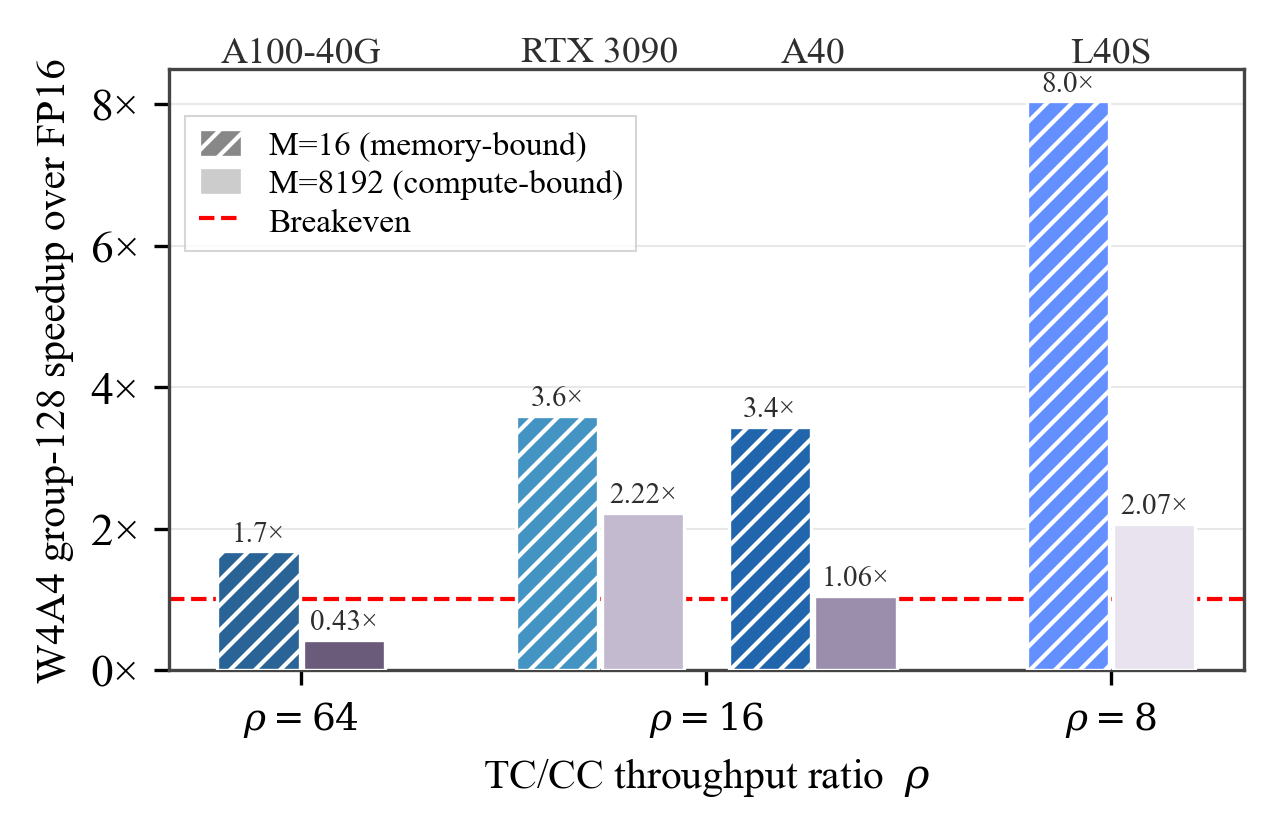} 
    \caption{The proposed W4A4-g128 GEMM kernel speedup over FP16 ($N{=}K{=}8192$) across GPUs with varying $\rho$. Higher $\rho$ consistently yields lower speedup; A100 ($\rho{=}64$) falls below break-even.}
    \label{fig:motivation_fig}
\end{wrapfigure}
 
The fundamental obstacle for W4A4 quantization is a coupled tension between accuracy and efficiency. 4-bit activation quantization is highly sensitive to outliers~\citep{smoothquant,Quarot}, and fine-grained group quantization is the key strategy to mitigate this accuracy loss~\citep{Chen25}. However, each group's partial sums require dequantization through INT32-to-FP32 conversion and scaling-factor multiplications on CUDA Cores (CC), whose throughput is slower than Tensor Cores~\citep{Qserve,liquidgemm}; as quantization granularity becomes finer, these operations multiply proportionally. 
This forms a vicious cycle: accuracy demands drive finer granularity,
while finer granularity amplifies the dequantization overhead,
ultimately making serial dequantization on CUDA Cores the primary performance bottleneck of W4A4 group-quantized GEMM kernels.

The central insight of this paper is that the severity of this bottleneck is not constant, but is largely governed by the intra-SM balance between Tensor Cores and CUDA Cores throughput, which we capture as $\rho = \mathcal{T}_{\mathrm{TC}} / \mathcal{T}_{\mathrm{CC}}$. 
This ratio varies significantly across GPUs, even among those sharing the same architecture and instruction set, implying that the feasibility of W4A4 group quantization should be assessed by the intra-SM compute balance captured by $\rho$, rather than by GPU generation alone.
Section~\ref{sec:motivation} substantiates this claim with controlled cross-architecture experiments revealing that the same W4A4 group-128 kernel yields up to $2.2\times$ speedup on a low-$\rho$ platform yet degrades to $0.43\times$ on a high-$\rho$ one, a gap of approximately $5\times$ with $\rho$ as a primary differentiator.

Building on these insights, this paper systematically quantifies how the intra-SM throughput ratio $\rho$, together with system-level resources such as memory bandwidth, determines the practical speedup ceiling of W4A4 group quantization across granularities and hardware platforms. 
While prior work has made hardware-specific granularity choices empirically (e.g., QServe~\citep{Qserve} selects per-channel quantization on A100 but per-group on L40S), these decisions lack a unifying quantitative basis; our analysis provides such a basis through $\rho$, enabling principled granularity selection for a given target GPU.

To isolate hardware effects from algorithmic artifacts, we adopt
a two-level controlled experimental design.
The first level is an intra-architecture comparison across three
Ampere GPUs (A100, RTX~3090, A40), where the ISA is held constant
while $\rho$ and memory bandwidth vary.
The second level introduces the Ada-based L40S to validate whether
the correlations established on Ampere generalize across GPU
generations.
Existing W4A4 kernels~\citep{COMET,ATOM} involve mixed-precision
Tensor Cores invocations that introduce significant confounding
variables; we therefore develop highly optimized \emph{pure} W4A4
group quantization kernels covering seven granularities, ensuring
that performance measurements faithfully reflect hardware
characteristics.
On the algorithm side,
pure INT4 arithmetic offers no mixed-precision fallback
to absorb quantization error.
We therefore combine activation smoothing~\citep{Quarot}
with block-wise distillation calibration~\citep{omniquant}
to suppress quantization noise,
and support both uniform group quantization (e.g., g128)
and a $\rho$-aware mixed-granularity strategy
that reserves fine-grained groups for accuracy-sensitive layers
while routing the majority through per-channel quantization,
jointly optimizing accuracy and kernel efficiency.

The main contributions of this work are as follows:

\textbf{Hardware characterization.}
    We conduct controlled kernel level benchmarks
    across four GPUs spanning two architecture generations
    (Ampere: A100, RTX~3090, A40; Ada: L40S)
    and identify the intra-SM \emph{Tensor-Core-to-CUDA-Core
    throughput ratio} $\rho$ as the primary hardware factor
    governing W4A4 group-quantization efficiency.
    The same W4A4-g128 kernel ranges from
    $2.0$--$2.5\times$ speedup ($\rho=16$, RTX~3090)
    to $0.43$--$0.47\times$ ($\rho=64$, A100),
    and cross-precision analysis shows that W4A4
    is substantially more $\rho$-sensitive
    than W4A16 and W4A8 group kernels
    due to serial dequantization on CUDA Cores
    within the GEMM main loop.

\textbf{Algorithm--kernel co-design.}
    Guided by the $\rho$ analysis,
    we co-design a quantization algorithm
    and an optimized INT4 GEMM kernel.
    On the algorithm side,
    we combine activation smoothing with block-wise
    distillation calibration,
    and propose a $\rho$-aware mixed-granularity strategy
    that assigns per-channel quantization to most layers
    while reserving $G{=}32$ for accuracy-sensitive ones.
    On the kernel side,
    we execute all matrix multiplications on INT4 Tensor Cores
    while dequantization is handled by CUDA Cores
    with software-pipelined scale loading,
    eliminating the mixed-precision Tensor Core fallbacks
    required by prior W4A4 kernels~\citep{ATOM,COMET}.
    The resulting W4A4-g128 achieves perplexity within 0.63
    of FP16 on Llama-2-70B
    and outperforms Atom-g128 by 4.0\%--4.4\%
    in average zero-shot accuracy on Llama-2.

    \textbf{System integration and evaluation.}
    We integrate the above into the APEX4,
    which adapts quantization granularity
    to the target GPU's $\rho$ from a single codebase.
    Deployed within vLLM,
    end-to-end serving benchmarks
    on Llama-2-7B and Qwen2.5-7B
    show that W4A4-g128 delivers consistent speedup
    on low-$\rho$ GPUs,
    up to $1.78\times$ on RTX~3090 and $1.66\times$ on L40S,
    while the mixed-granularity mode
    reaches $2.09\times$ on A40
    and recovers A100 to $1.20$--$1.40\times$
    at batch sizes $\geq$64,
    demonstrating that $\rho$-aware configuration
    makes W4A4 a practical serving primitive
    across architecturally diverse GPU deployments.

\section{Dissecting the W4A4 Performance Gap}
\label{sec:motivation}

\subsection{Faster Hardware, Slower Kernel}
\label{sec:observation}
 \begin{table}[htbp]
\centering
\footnotesize                    
\setlength{\tabcolsep}{3pt}      

\caption{Evaluated GPU specifications. $\rho$ denotes the per-SM 
INT4 Tensor Core to FP32 CUDA Core throughput ratio, the key 
predictor of W4A4 kernel speedup.}

\begin{tabular}{lcccccccccc}
\toprule
\textbf{GPU} & \textbf{Arch.} & \textbf{SMs}
& \textbf{TC/SM}
& \textbf{CC/SM}
& \textbf{INT4 TC}
& \textbf{INT4 TC} & \textbf{FP32 CC} & \textbf{$\rho$}
& \textbf{Bandwidth} & \textbf{L2 Cache} \\
 & & & & & \textbf{OPs/cycle} & \textbf{dense (TOPS)} & \textbf{(TFLOPS)} & & \textbf{(GB/s)} & \textbf{(MB)} \\
\midrule
A100-40G & Ampere & 108 & 4 & 64  & 2,048 & 1,248 & 19.5 & 64 & 1,555 & 40 \\
RTX 3090 & Ampere & 82  & 4 & 128 & 1,024 & 568   & 35.6 & 16 & 936   & 6  \\
A40      & Ampere & 84  & 4 & 128 & 1,024 & 599   & 37.4 & 16 & 696   & 6  \\
L40S     & Ada    & 142 & 4 & 128 & 512   & 733   & 91.6 & 8  & 864   & 96 \\
\bottomrule
\end{tabular}
\label{tab:gpu_comparison}
\end{table}
To understand why W4A4 group quantization performs so differently across hardware, we deploy the same proposed W4A4-g128 kernel on four GPUs (see Table \ref{tab:gpu_comparison}) and measure its speedup over FP16 at two representative matrix dimensions: $M = 16$ (memory-bound) and $M = 8192$ (compute-bound). 

Fig.~\ref{fig:motivation_fig} reveals the counter-intuitive result: among the four platforms, the A100-40G has the highest INT4 Tensor Cores throughput (1248 TOPS) and the largest memory bandwidth (1555 GB/s), yet performs the worst in both scenarios. In the memory-bound scenario, the A100 achieves only $1.7 \times$ speedup, well below the $8.0 \times$ of the L40S and $3.6 \times$ of the RTX 3090; in the compute-bound scenario, the A100's speedup drops further to $0.43 \times$, offering no acceleration.

This result cannot be explained by Tensor Cores throughput or memory bandwidth, because the A100-40G leads on both metrics. The W4A4 group quantization kernel does not rely solely on Tensor Cores: the dequantization operations within its main loop execute on CUDA Cores at FP32 precision. When we instead examine the throughput ratio $\rho$ between Tensor Cores and CUDA Cores (lower values indicate relatively more abundant CUDA Core capacity), the ranking reverses: $\rho$ is 8 for the L40S, 16 for the RTX 3090 and A40, and 64 for the A100. The A100, with the highest ratio, is precisely the platform that falls below the breakeven line in the compute-bound scenario.
 
\subsection{The Primary Bottleneck Factor}
\label{sec:cuda_core_density}

A natural follow-up question is: why does the throughput ratio $\rho$ (between Tensor Cores and CUDA Cores) drive such performance gaps across GPUs?

\begin{wrapfigure}{r}{0.5\textwidth}
    \centering
    \includegraphics[width=0.5\textwidth]{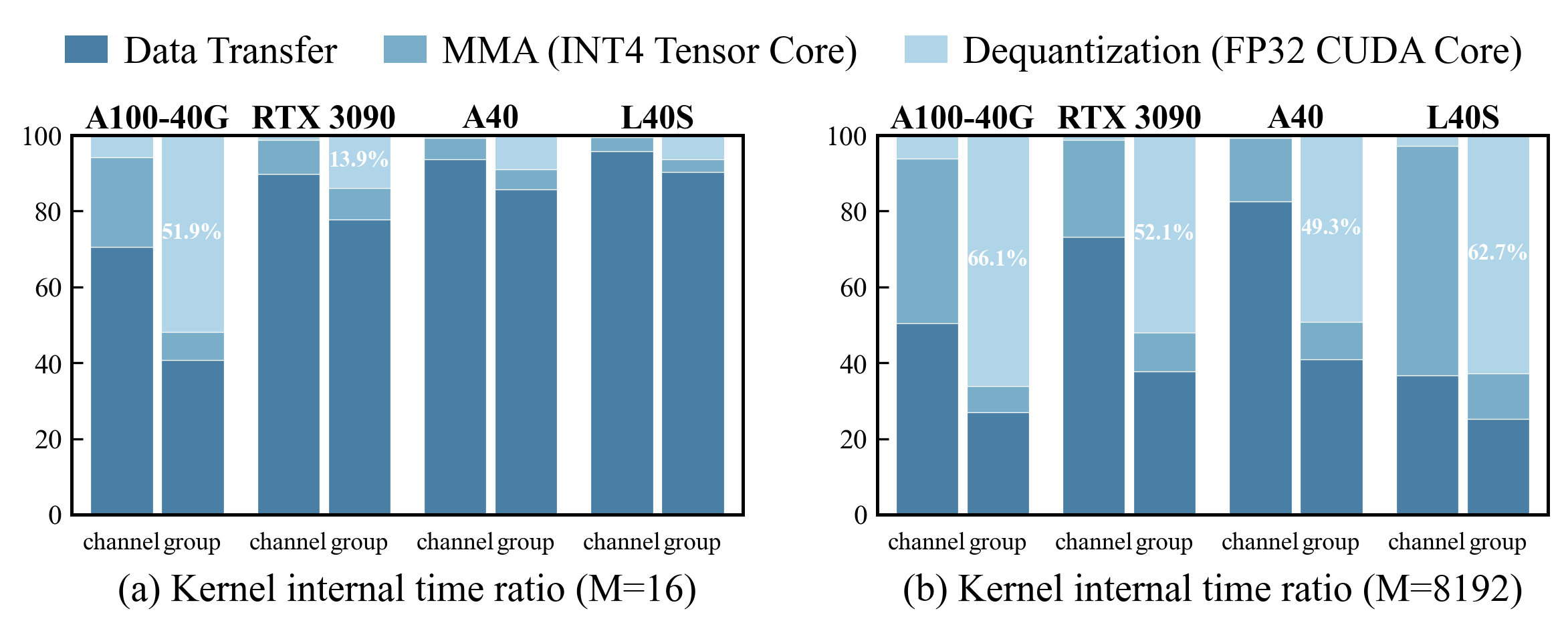} 
    \caption{Kernel-internal time ratio of W4A4 channel and group32 quantization across four GPUs.}
    \label{fig:motivation_kernel_internal}
\end{wrapfigure}

For an $M \times N \times K$ matrix multiplication, per-channel quantization shares a single scaling factor across the entire $K$ dimension, allowing dequantization to be completed in a post-processing stage outside the main loop. Group quantization, in contrast, partitions $K$ into groups of size $G$, each with an independent scaling factor, requiring dequantization to be performed iteratively within the main loop for a total of $K/G$ times (see Section~\ref{sec:Symmetric_Formulation}).
As shown in Fig.~\ref{fig:motivation_kernel_internal},
on every platform the dequantization fraction
rises substantially when switching
from per-channel to group32 quantization.
For example, on the A100 at $M{=}8192$,
the fraction increases from 6.2\% to 66.1\%,
confirming that the $K/G$-fold
in-loop dequantization is the dominant source
of overhead in group quantization kernel.

However, all four platforms execute the same number of dequantization operations, so the $K/G$-fold increase cannot explain the cross-platform performance disparity. The key lies in the relative cost of each dequantization operation: dequantization executes on FP32 CUDA Cores, while MMA executes on INT4 Tensor Cores. On data center GPUs such as the A100, the peak throughput of FP32 CUDA Cores is roughly 2\% of their INT4 Tensor Cores~\citep{Qserve}. Moreover, dequantization depends on the output of the preceding MMA, creating a data dependency that prevents the two from being overlapped via pipelining.

This is precisely what $\rho$ captures: a higher $\rho$ means each dequantization is more expensive relative to MMA. As $\rho$ grows, the cumulative overhead of $K/G$ dequantization steps increasingly dominates kernel execution time.
Our profiling data corroborate the above analysis. As shown in Fig.~\ref{fig:motivation_kernel_internal}, the A100, which has the highest
$\rho$ (64) among the four platforms,
consistently exhibits a larger dequantization
fraction than all other GPUs
in both the compute-bound ($M{=}8192$, 66.1\%)
and memory-bound ($M{=}16$, 51.9\%) scenarios.

The above analysis indicates that the CUDA Cores capacity should not fall too far behind the Tensor Cores capacity to achieve good performance for W4A4 group quantization.
An intuitive approach would be to use the ratio of Tensor Cores to CUDA Core counts to gauge this imbalance. Core counts do provide a rough reference: the three platforms with 128 CC/SM (RTX 3090, A40, L40S) generally outperform the A100 with 64 CC/SM. However, actual throughput depends not only on core counts but also on clock frequency and operations per cycle:
\begin{equation}
  \mathcal{T} = Number_{\text{cores}} \times f_{\text{clk}} \times \text{OPs/cycle}
\end{equation}
Since the TC and CC within the same GPU share a common clock frequency, the frequency term cancels in $\rho$:
\begin{equation}
  \rho = \frac{N_{\mathrm{TC}} \times \text{OPs}_{\mathrm{TC}}}{N_{\mathrm{CC}} \times \text{OPs}_{\mathrm{CC}}}
\end{equation}
That is, $\rho$ is determined by the ratio of core counts and the ratio of per-cycle operations, and core counts alone cannot capture the actual disparity in computational capacity. Furthermore, the per-cycle INT4 operations of Tensor Cores vary considerably across architectures (see Table~\ref{tab:gpu_comparison}), enabling $\rho$ to further reflect the computational imbalance across different GPUs. This is why we adopt a throughput ratio rather than a core count ratio.

In summary, the performance bottleneck of group quantization is jointly determined by two factors: the algorithmic dequantization frequency $K/G$, and the hardware-dependent relative dequantization cost captured by $\rho$. 
The former is governed by quantization granularity and is identical across platforms; the latter varies with hardware and is the primary source of cross-platform performance divergence. Secondary factors such as memory bandwidth and L2 Cache size also influence absolute performance, 
but our cross-platform experiments indicate that $\rho$ is the leading predictor of group quantization speedup. 
Section~\ref{sec:experiment} provides systematic validation.

This conclusion also explains why we focus on specific hardware. On
Ampere and Ada, INT4 is computed natively on the Tensor Cores, while the
dequantization of group quantization runs as a separate step on the CUDA
Cores, which is where the bottleneck arises. Newer architectures no
longer work this way: on Hopper, INT4 computation degrades to CUDA Core
operations~\citep{luo2024benchmarking} and cannot attain Tensor Core throughput; Blackwell removes
INT4 outright in favor of floating-point FP4 and folds block scaling into
the MMA instruction~\citep{nvidia_blackwell_whitepaper}, so that FP4 needs no separate dequantization step.
In short, neither generation follows the Tensor-Core INT4 path. The
bottleneck we characterize is therefore specific to the Ampere and Ada
generations, which remain widely used and are the focus of this work.


\section{Pure W4A4 Quantization Algorithm Design}
\label{sec:algorithm}

The preceding analysis identifies \emph{when} W4A4 group quantization
can be efficient; this section addresses \emph{whether} it can also
be accurate. Without mixed-precision fallback, we develop three
components: activation smoothing
(Section~\ref{sec:smoothing}), hardware-aware multi-granularity
quantization (Section~\ref{sec:quant_strategy}), and block-wise
knowledge distillation (Section~\ref{sec:ptq}).


\subsection{Hadamard-Based Activation Smoothing}
\label{sec:smoothing}

\begin{figure}
    \centering
    \includegraphics[width=1\textwidth]{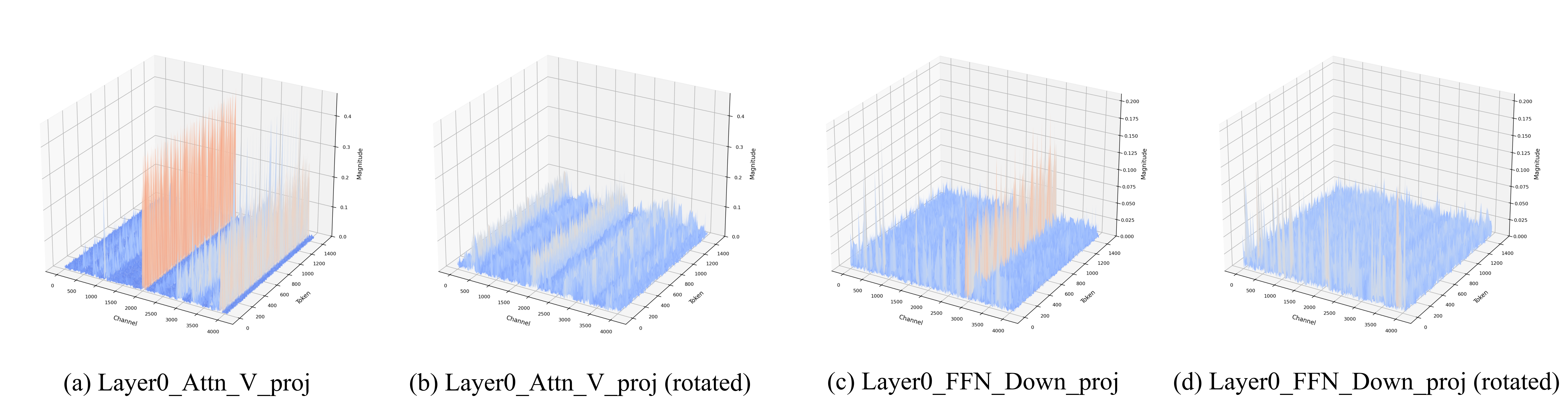} 
    \caption{Effect of Hadamard-based activation smoothing. }
    \label{fig:hadamard}
\end{figure}

To suppress the sparse high-magnitude activation outliers that dominate 4-bit quantization error, we apply offline orthogonal rotations following QuaRot~\citep{Quarot}, redistributing outlier energy evenly across all channels. Fig.~\ref{fig:hadamard} visualizes the activation distribution of Attn\_V\_proj and FFN\_Down\_proj
before and after the rotation, showing a significant reduction 
in outlier magnitude.

Concretely, let $Q$ be a randomized Hadamard matrix $\tilde{H} = H \cdot \text{diag}(s)$ satisfying $QQ^T = I$. Each pair of adjacent linear layers absorbs $Q$ (output side) and $Q^T$ (input side). After fusing RMSNorm parameters into the weights, the rotations are:
\begin{align}
W_{\text{embed}} &\leftarrow W_{\text{embed}}\, Q, \quad W_{\text{head}} \leftarrow W_{\text{head}}\, Q, \\
W_{qkv} &\leftarrow W_{qkv}\, Q, \quad W_o \leftarrow Q^T W_o, \\
W_{\text{up}}, W_{\text{gate}} &\leftarrow W_{\text{up}}\, Q, \; W_{\text{gate}}\, Q, \quad W_{\text{down}} \leftarrow Q^T W_{\text{down}}.
\end{align}
The paired $Q$ and $Q^T$ cancel at each layer boundary, so intermediate activations (q, k, v, MLP hidden states) remain in the original unrotated space. To further address outlier concentrations \emph{within} individual attention heads, we apply exact Hadamard transformations to value-output projection pairs at the per-head granularity:
\begin{equation}
W_v \leftarrow W_v\, H_{\text{head}}, \quad W_o \leftarrow H_{\text{head}}^T W_o.
\end{equation}

Unlike the full QuaRot pipeline, all transformations are restricted to offline weight preprocessing, avoiding runtime CUDA-core overhead (Section~\ref{sec:cuda_core_density}).


\subsection{W4A4 Group Quantization Strategy}
\label{sec:quant_strategy}
With activation outliers suppressed, the remaining design choice is quantization granularity. Finer group sizes improve local distribution modeling but proportionally increase the dequantization workload on CUDA Cores, so the strategy must balance accuracy against the $\rho$-dependent efficiency cost.
\subsubsection{Symmetric Formulation}
\label{sec:Symmetric_Formulation}
We adopt symmetric quantization for both weights and activations, eliminating zero-point parameters and reducing dequantization to a single scaling multiplication per side---directly benefiting kernel efficiency by removing the integer subtractions required by asymmetric schemes.

For a matrix $X$ (weight or activation) partitioned along the reduction dimension $K$ into groups of size $G$, each group $X_g$ is independently quantized as:
\begin{equation}
S_g = \frac{\max(|X_g|)}{2^{b-1} - 1}, \quad
X_g^{\text{q}} = \text{clamp}\!\left(\left\lfloor\frac{X_g}{S_g}\right\rceil,\; -2^{b-1},\; 2^{b-1}{-}1\right),
\end{equation}
where $b$ is the bit-width and $S_g$ is the group-specific scaling factor
(weights quantized offline, activations dynamically at inference).
When both operands are group-quantized,
the $M \times N \times K$ matrix multiplication
decomposes into $K/G$ partial products:
\begin{equation}
\label{eq:group_decomposition}
C = \sum_{g=0}^{K/G-1}
  \left(A_g^{\mathrm{q}} \cdot W_g^{\mathrm{q}}\right)
  \odot
  \left(S_g^{a} \cdot {S_g^{w}}^{\!\top}\right),
\end{equation}
where each $A_g^{\mathrm{q}} \cdot W_g^{\mathrm{q}}$ produces $M \times N$ INT32 partial sums on Tensor Cores,
and $\odot$ applies the outer product of activation and weight scaling factors via element-wise FP32 multiplication on CUDA Cores.
This dequantization executes once per group,
introducing $K/G$ serial CUDA Core operations
into the GEMM main loop (Section~\ref{sec:cuda_core_density}).

\subsubsection{Multi-Granularity Configuration}
\label{sec:multi_granularity}
\begin{figure}[!htb]
    \centering
    \includegraphics[width=\textwidth]{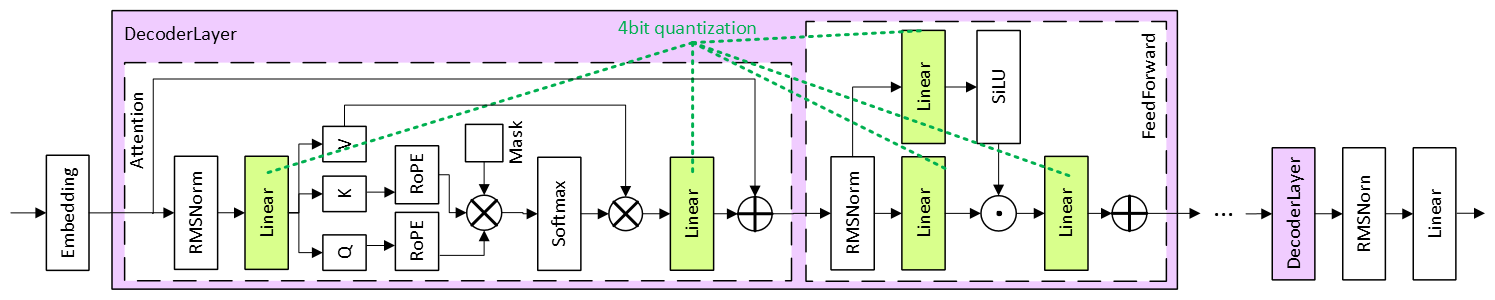} 
    \caption{Overview of transformer architecture with W4A4 quantization deployment. The linear layers subject to W4A4 quantization are highlighted, encompassing query ($W_q$), key ($W_k$), and value ($W_v$) projections within the multi-head attention mechanism, the attention output projection ($W_o$), and feed-forward network linear transformations including up-projection ($W_{\text{up}}$), gate-projection ($W_{\text{gate}}$), and down-projection ($W_{\text{down}}$).}
    \label{fig:model_structure}
\end{figure}

Not all layers exhibit the same sensitivity to quantization granularity. We exploit this heterogeneity to concentrate fine-grained quantization where the accuracy return is highest. Weight and activation group sizes are kept equal ($G_w = G_a = G$) within each layer. Based on empirical layer-sensitivity analysis, we propose two configurations.

Uniform $G{=}128$ applies a single group size across all linear layers, serving as a baseline with moderate accuracy--efficiency trade-offs.

Mixed granularity differentiates by layer function:
 $W_{\text{down\_proj}}$ and $W_{\text{v\_proj}}$: $G{=}32$. These layers are most sensitive: down-projections amplify per-element errors across all output dimensions, and value projections propagate distortions through the softmax nonlinearity.
All other linear layers: per-channel quantization ($G{=}K$), incurring minimal dequantization overhead.

This co-design allocates the majority of layers to the most hardware-efficient granularity while reserving fine-grained protection for the two most sensitive layer types. Fig.~\ref{fig:model_structure} illustrates both configurations.
The practical impact varies with the target GPU's $\rho$ and is quantitatively evaluated in Section~\ref{sec:experiment}. 

\subsection{Block-wise Knowledge Distillation}
\label{sec:ptq}
The preceding two components reduce but do not eliminate quantization error. Under pure W4A4 precision, conventional PTQ methods such as GPTQ~\citep{GPTQ} and AWQ~\citep{AWQ} prove insufficient, as the discrete 4-bit weight space is too coarse for their optimization strategies. Inspired by OmniQuant~\citep{omniquant}, we jointly optimize both scaling factors $\{S_g\}$ and quantized weights $\{W_g^{\text{q}}\}$ through block-wise knowledge distillation.

The optimization follows a greedy block-by-block strategy (Algorithm~\ref{alg:block_distillation}). For each transformer block $\mathcal{B}_i$, we minimize the cosine distance between the full-precision and quantized block outputs:
\begin{equation}
\min_{\{S_g\},\, \{W_g^{\text{q}}\}} \;\mathcal{L}_i = 1 - \frac{\mathcal{B}_i(X_i^{\text{q}};\, \Theta_i^{\text{FP}}) \cdot \mathcal{B}_i(X_i^{\text{q}};\, \Theta_i^{\text{Q}})}{|\mathcal{B}_i(X_i^{\text{q}};\, \Theta_i^{\text{FP}})| \;\, |\mathcal{B}_i(X_i^{\text{q}};\, \Theta_i^{\text{Q}})|}\,,
\end{equation}

where $X_i^{\text{q}}$ is the output of the previously optimized block $\mathcal{B}_{i-1}$. This greedy cascade ensures that each block's optimization accounts for cumulative quantization effects from all preceding layers. Cosine similarity (chosen for scale invariance over MSE) provides stable gradients across layers with varying output norms. Gradients through the discrete quantization are approximated via the straight-through estimator (STE), with weights re-quantized after each update.

With accuracy addressed, we turn to the system side: designing GPU kernels that translate the low-bit advantage into measured speedup across architectures with different $\rho$.


\begin{algorithm}[t]
\caption{Greedy Block-wise Knowledge Distillation}
\label{alg:block_distillation}
\begin{algorithmic}
\State \textbf{Input:} Full-precision model $\{\mathcal{B}_1, \dots, \mathcal{B}_L\}$, calibration data $X_0$
\State Initialize $X_0^{\text{q}} \leftarrow X_0$
\For{$i = 1$ to $L$}
  \State Initialize $\{S_g\}$, $\{W_g^{\text{q}}\}$ for block $\mathcal{B}_i$
  \For{$t = 1$ to $T_{\text{calib}}$}
    \State $\hat{Y} \leftarrow \mathcal{B}_i(X_i^{\text{q}};\; \{S_g\}, \{W_g^{\text{q}}\})$ \Comment{Quantized forward}
    \State $Y \leftarrow \mathcal{B}_i(X_i^{\text{q}};\; W_i^{\text{FP}})$ \Comment{Full-precision forward}
    \State $\mathcal{L}_i \leftarrow 1 - \cos(\hat{Y},\, Y)$
    \State Update $S_g$, $W_g^{\text{real}}$ via STE
    \State $W_g^{\text{q}} \leftarrow \text{SymQuant}(W_g^{\text{real}},\, S_g)$
  \EndFor
  \State $X_{i+1}^{\text{q}} \leftarrow \mathcal{B}_i(X_i^{\text{q}};\; \{S_g\}, \{W_g^{\text{q}}\})$ \Comment{Propagate quantized output}
\EndFor
\State \textbf{where} $\text{SymQuant}(W, S) = \text{clip}\!\left(\left\lfloor W / S \right\rceil,\, -2^{b-1},\, 2^{b-1}-1\right)$
\end{algorithmic}
\end{algorithm}

\section{W4A4 Kernel Design} 
\label{sec:kernel}

\subsection{Design Overview}
We propose a unified fine-grained W4A4 quantization kernel architecture that includes a weight-matrix-centric striped partitioning strategy for uniform distribution of computational resources, a four-stage asynchronous pipeline that hides data loading latencies, and a unified data preprocessing workflow that improves data loading efficiency.
Specifically, the kernel architecture supports both channel quantization and group configurations (32, 64, 128, 256, 512, 1024). 
Due to different optimization needs for various quantization granularities, we use dual-kernel designs for channel-wise and group-wise quantization. Both kernels employ symmetric quantization to avoid the additional computational overhead of zero-point offsets present in asymmetric quantization.
Compared to mixed-precision approaches, our pure 4-bit design unifies data precision and fully leverages the computational capabilities of INT4 Tensor Cores.

\begin{figure*}[!htb]
    \centering
    \includegraphics[width=\textwidth]{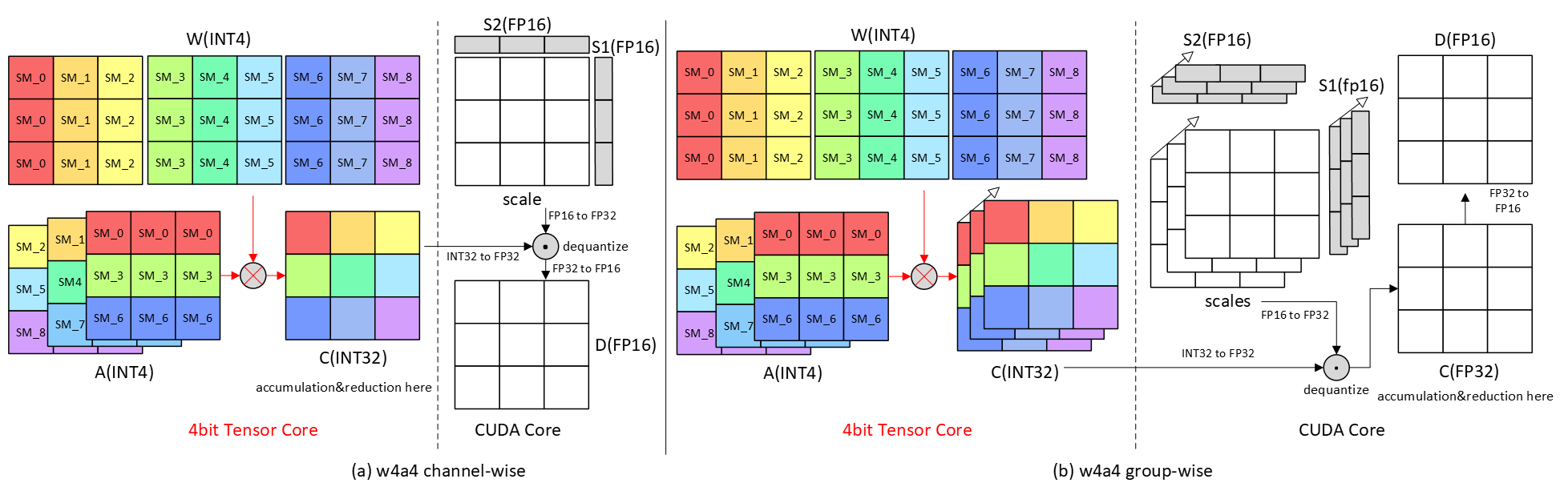} 
    \caption{W4A4 channel and group quantization principal overview. Subfigures (a) and (b) respectively demonstrate channel and group quantization in our scheme at the SM resource allocation and tile-level computation. The INT4 matrix sizes are 192×768×384 (M×N×K), with M\_tile=3, N\_tile=3, K\_tile=3.}
    \label{fig:w4a4_channel_group_overview}
\end{figure*}

\begin{figure*}[!htb]
    \centering
    \includegraphics[width=\textwidth]{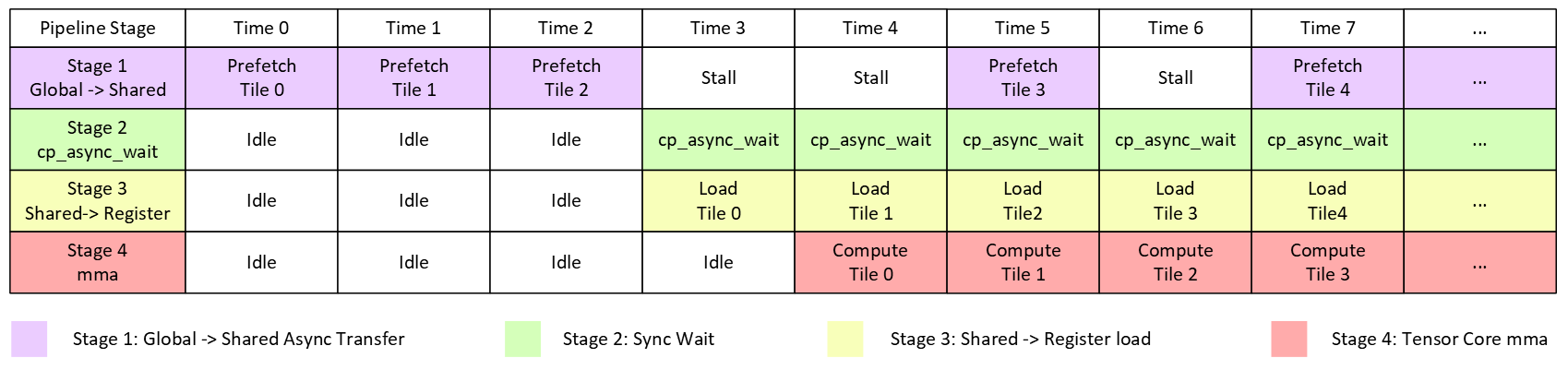} 
    \caption{Four-stage asynchronous pipeline timing diagram.}
    \label{fig:pipeline}
\end{figure*}

\subsection{Kernel Principles}
\label{sec:kernel_principles}

Matrix multiplication adopts a weight-matrix-centric striped partitioning scheme, where each block computes fixed-size tiles to ensure uniform distribution of computational resources. We use Tensor Core's m16n8k32.s4.s4.s32 instruction, and based on this constraint, set block computation tile sizes: Activation (A) as $64 \times 128 (M \times K)$, Weight (W) as $128 \times 256 (K \times N)$.

As shown in Fig.~\ref{fig:w4a4_channel_group_overview}, assuming a GPU contains 10 SMs and input matrices A and W are both 3×3 tile blocks, 
\begin{equation}
SM\_iterations = \left\lceil \frac{M\_tile \times N\_tile \times K\_tile}{Number\_of\_SMs} \right\rceil.
\end{equation}
When tiles in the same column are computed by multiple SMs, inter-SM global reduction is performed to ensure accumulation correctness.

Dequantization strategy is the core difference between channel and group kernels, directly affecting computational efficiency, memory access patterns, and CUDA Core resource allocation.
The channel quantization kernel employs delayed dequantization. As shown in Fig.~\ref{fig:w4a4_channel_group_overview} (a), the delayed dequantization strategy completes all matrix block MMA computations and accumulates them to $C$, then performs dequantization once at the end.

Group quantization kernel uses immediate dequantization.
As shown in Fig.~\ref{fig:w4a4_channel_group_overview}(b), since different groups correspond to different scaling factors, dequantization cannot be deferred. After each SM completes tile matrix multiplication and obtains tile $C$, it immediately scales $C$ using pre-loaded $S1$ and $S2$. Global accumulation is performed after all tiles are dequantized.
Immediate dequantization introduces frequent scale factor loads. To mitigate this, we pipeline the loading of scale factors and design a unified dispatch mechanism that adapts to varying group sizes, ensuring correct scale-to-data alignment across configurations.

\begin{figure}
    \centering
    \includegraphics[width=1\textwidth]{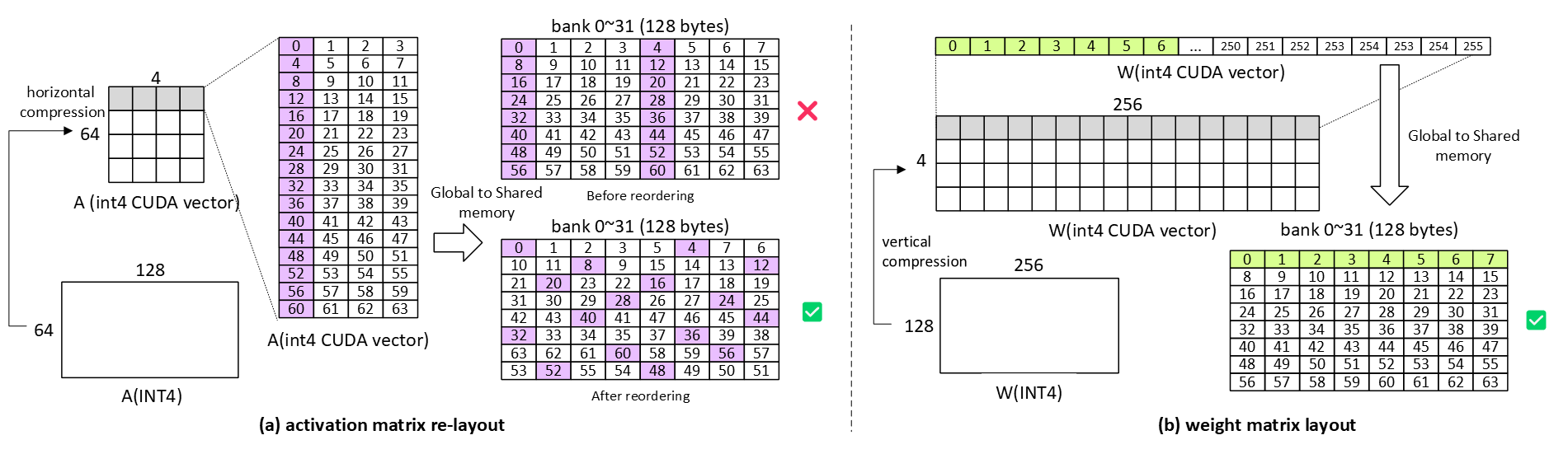} 
    \caption{Activation matrix and weight matrix data preprocessing and bank conflict avoidance principle.}
    \label{fig:bank_conflict}
\end{figure}

\begin{wrapfigure}{r}{0.5\textwidth}
    \centering
    \includegraphics[width=0.5\textwidth]{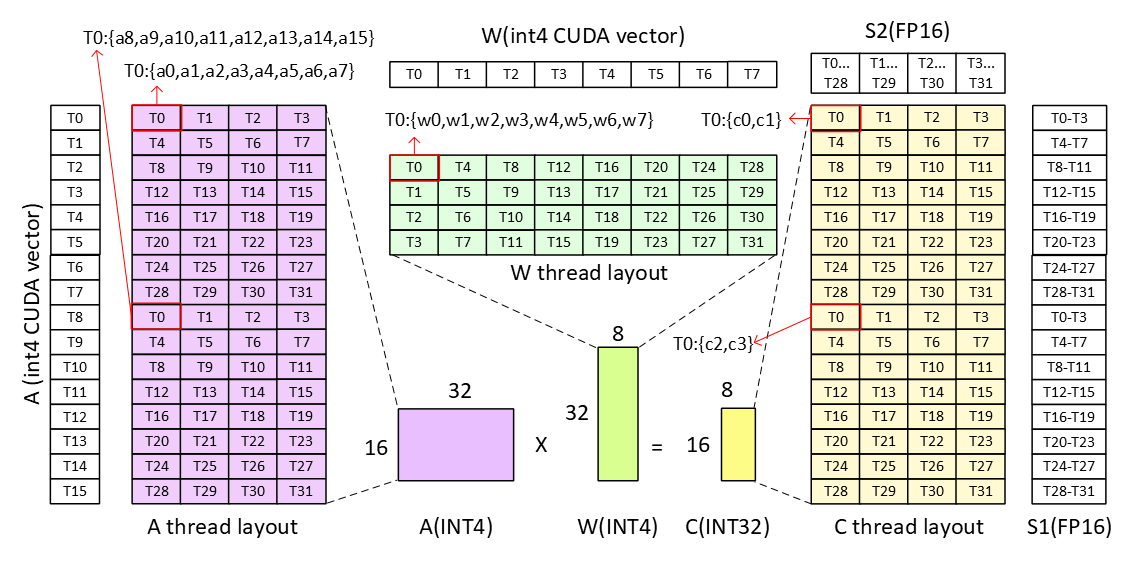} 
    \caption{Thread layouts of activation matrix, weight matrix, and result matrix C on the minimal instruction-level tile, as well as the thread layouts when loading activation and weight matrix data using ldmatrix instructions, and the thread layouts when loading S1 and S2.}
    \label{fig:ldmatrix}
\end{wrapfigure}
\subsection{Four-stage Asynchronous Pipeline}
\label{sec:pipeline}
 
The pipeline contains four key stages.
(1) Global→Shared asynchronous transfer stage: uses cp.async.cg.shared.global instructions for 16-byte aligned transfers, asynchronously prefetching activation matrices, weight matrices, and group scale factors to hide global memory access latency; 
(2) Synchronization wait stage: uses cp\_async\_wait<stages-2>() to ensure completion of global to shared memory loading; 
(3) Shared→Register loading stage: uses ldmatrix instructions to transfer activation and weight data from shared memory to corresponding register positions, along with group scale factor movement; 
(4) MMA computation stage, where group quantization kernels immediately execute dequantization computation, while channel quantization kernels maintain INT32 accumulation with delayed dequantization. The channel scale factors will be loaded after the 4-stage pipeline.
Once the pipeline reaches steady state, all four stages operate concurrently on different data blocks, overlapping data loading with computation.
Details are shown in Fig.~\ref{fig:pipeline}

\subsection{Data Management Strategy}
\label{sec:data_management}

Data preprocessing and memory management are key foundations for efficient kernel execution, involving the design of activation matrices, weight matrices, and scale factors.

Activation Matrix Processing: 
Taking tiles processed by each block as an example, FP16 activation matrix A is quantized online to pure 4-bit and horizontally compressed to a 64×4 int4 CUDA vector layout. 
As shown in Fig.~\ref{fig:bank_conflict} (a), if a 16 $\times$ 4 int4 vector matrix block is loaded into shared memory, 8 threads in the same warp simultaneously reading the first column's 8 data points would cause 8 memory conflicts. 
We pre-calculate all transform addresses at compile time to ensure that only 2 threads access the same bank simultaneously.
Activation matrix loading from shared memory to registers uses ldmatrix instructions. The minimum tile at instruction level is 16$\times$32 (4bit), which can be horizontally compressed into 16$\times$1 int4 CUDA vector data, so ldmatrix instructions can directly load these 16 data addresses(refer to A thread layout in Fig.~\ref{fig:ldmatrix} for details).

Weight Matrix Processing: 
Taking tiles processed by each block as an example, weight matrices are offline quantized to 4-bit and vertically compressed into 4×256 int4 CUDA vector matrices. 
As shown in Fig.~\ref{fig:bank_conflict} (b), assuming the first row of data is loaded into shared memory, the 8 data points that the same warp needs to read simultaneously are exactly in the first row of shared memory, naturally avoiding queuing conflicts while maintaining address continuity during data loading. 
Similarly, we use ldmatrix instructions to load weight matrices. As shown in Fig.~\ref{fig:ldmatrix} , the minimum tile at instruction level after vertical compression changes from 32×8 4bit to 1$\times$8 int4 CUDA vectors, directly using ldmatrix instructions to load these 8 vector data addresses.

Scale Factor Management:
Both activation scale factor ($S1$) and weight scale factor ($S2$) are in FP16 format.Before being loaded into registers by threads according to specific layouts, $S1$ and $S2$ first undergo data type conversion (FP16 to FP32).
At the minimal instruction-level tile, $S1$ of length 16 and $S2$ of length 8 need to be precisely allocated according to the thread layout requirements of matrix $C$, ensuring efficient access during dequantization. For example, as shown in Fig.~\ref{fig:ldmatrix}, threads 0-3 need to load the first and eighth data elements of S1; S2 needs to load data 0 and 1 to threads 0, 4, 8, 12...28.

\section{Evaluation}
\label{sec:experiment}
\subsection{Experimental Setup}

Following the multi-granularity strategy described in Section~\ref{sec:multi_granularity}, we evaluate two quantization configurations: W4A4-g128 and W4A4-mix.
This dual-configuration span the accuracy–efficiency trade-off space and allow us to isolate the effect of quantization granularity on both model accuracy and hardware efficiency across different GPU architectures.

The block-wise full-parameter optimization employs the Adam optimizer with learning rate 1e-5 and standard momentum parameters ($\beta_1=0.9$, $\beta_2=0.999$). Calibration is performed using 256 samples from the WikiText-2 training set over 10 epochs for each transformer block, providing sufficient adaptation while maintaining computational efficiency for large-scale models.

Hardware efficiency experiments follow a two-level design: intra-architecture comparisons across three Ampere GPUs (A100, A40, RTX~3090) that share the same ISA but span a wide range of $\rho$ and memory bandwidth, and a cross-generation check using the Ada-based L40S to verify generalizability.
We do not include Turing in our evaluation: although Turing also has
native INT4 Tensor Cores, our kernel pipeline relies on the
\texttt{cp.async} asynchronous-copy instruction introduced with Ampere,
which Turing does not support.
All experiments are conducted using CUDA 12.8, PyTorch 2.10, Transformers 4.57.6, and vLLM 0.19.0 to ensure reproducibility and compatibility with modern deployment environments.
Model quantization and evaluation follow fair comparison principles, with identical random seeds and evaluation protocols applied consistently across all configurations and hardware platforms.

\subsection{Model Accuracy Evaluation}

\textbf{Benchmarks.}
We evaluate our W4A4 APEX4 quantization approach on Llama-1~\citep{touvron2023llama}, Llama-2~\citep{touvron2023llama2}, Llama-3 families, and Qwen2.5~\citep{qwen2025qwen25technicalreport} (7B, 14B, 32B).
Perplexity is measured on WikiText-2~\citep{merity2016pointer}
with sequence length 2048.
Zero-shot accuracy is evaluated on five tasks
(PIQA~\citep{bisk2020piqa}, ARC-E/C~\citep{clark2018think},
HellaSwag~\citep{zellers2019hellaswag}, WinoGrande~\citep{sakaguchi2021winogrande})
via lm-evaluation-harness~\citep{eval-harness}.

\textbf{Baselines.}
We compare against representative post-training quantization methods:
SmoothQuant~\citep{smoothquant} (W8A8),
GPTQ~\citep{GPTQ} and AWQ~\citep{AWQ} (W4A16),
QoQ~\citep{Qserve} and QQQ~\citep{QQQ} (W4A8),
Atom~\citep{ATOM} and COMET~\citep{COMET} (W4A4 mixed-precision).
W4A16-Marlin~\citep{marlin} is included as a kernel and end-to-end performance baseline only.

\begin{table}[htbp]
\centering
\caption{WikiText2 perplexity ($\downarrow$) across quantization methods and model families.
W4Ax refers to 4-bit weights with mixed 4/8-bit activation precision. "–" marks results unavailable from the original publications.
}
\label{tab:ppl_wikitext2}
\begin{tabular}{ll ccc ccc c ccc}
\toprule
 & & \multicolumn{3}{c}{\textbf{Llama-1}} & \multicolumn{3}{c}{\textbf{Llama-2}} & \textbf{Llama-3} & \multicolumn{3}{c}{\textbf{Qwen2.5}} \\
\cmidrule(lr){3-5} \cmidrule(lr){6-8} \cmidrule(lr){9-9} \cmidrule(lr){10-12}
\textbf{Precision} & \textbf{Method} & 7B & 13B & 30B & 7B & 13B & 70B & 8B & 7B & 14B & 32B \\
\midrule
FP16 & -- & 5.68 & 5.09 & 4.10 & 5.12 & 4.57 & 3.12 & 6.14 & 6.84 & 5.29 & 5.01 \\
\midrule
W8A8 & SmoothQuant & 5.78 & 5.19 & 4.23 & 5.54 & 4.95 & 3.36 & 6.28 & {--} & {--} & {--} \\
\midrule
\multirow{2}{*}{W4A16}
 & GPTQ-g128         & 5.83 & 5.20 & 4.22 & 5.63 & 4.99 & 3.43 & 6.56 & {--} & {--} & {--} \\
 & AWQ-g128          & 5.78 & 5.19 & 4.21 & 5.60 & 4.97 & 3.41 & 6.54 & {--} & {--} & {--} \\
\midrule
\multirow{4}{*}{W4A8}
 & QoQ               & 5.93 & 5.28 & 4.34 & 5.75 & 5.12 & 3.52 & 6.89 & {--} & {--} & {--} \\
 & QoQ-g128          & 5.89 & 5.25 & 4.28 & 5.70 & 5.08 & 3.47 & 6.76 & {--} & {--} & {--} \\
 & QQQ               & 6.19 & 5.43 & 4.61 & 5.95 & 5.21 & 3.68 & 7.41 & {--} & {--} & {--} \\
 & QQQ-g128          & 5.87 & 5.24 & 4.30 & 5.71 & 5.01 & 3.50 & 6.64 & {--} & {--} & {--} \\
\midrule
\multirow{3}{*}{W4Ax}
 & Comet:W4AxKV4     & 5.95 & 5.32 & 4.31 & 5.73 & 5.19 & 3.56 & 6.91 & {--} & {--} & {--} \\
 & Comet:W4Ax        & 5.88 & 5.29 & 4.27 & 5.71 & 5.10 & 3.48 & 6.88 & {--} & {--} & {--} \\
 & Atom-g128     & 6.25 & 5.52 & 4.61 & 6.12 & 5.31 & 3.73 & 7.76 & {--} & {--} & {--} \\
\midrule
\multirow{2}{*}{W4A4}
 & \textbf{APEX4}-g128 & 6.17 & 5.47 & 4.54 & 6.09 & 5.28 & 3.75 & 7.70 & 7.87 & 6.94 & 6.08 \\
 & \textbf{APEX4}-mix     & 6.34 & 5.56 & 4.76 & 6.10 & 5.47 & 3.87 & 7.85 & 8.09 & 7.24 & 6.20 \\
\bottomrule
\end{tabular}
\end{table}

\begin{table}[htbp]
\centering
\small
\caption{Zero-shot accuracy (\%$\uparrow$) across Llama-2 models with various quantization methods. W4Ax refers to 4-bit weights with mixed 4/8-bit activation precision.}
\label{tab:zeroshot_performance}
\resizebox{\columnwidth}{!}{
\footnotesize  
\begin{tabular}{llcccccc}
\toprule
\textbf{Type} & \textbf{Method} & \textbf{PIQA} & \textbf{ARC-e} & \textbf{ARC-c} & \textbf{HellaSwag} & \textbf{WinoGrande} & \textbf{Avg.} \\
\midrule
\multicolumn{8}{l}{\textbf{Llama-2-7B}} \\
\cmidrule(lr){1-8}
 FP16 & --                    & 79.05 & 74.58 & 46.25 & 76.05 & 68.98 & 68.98 \\
\cmidrule(lr){1-8}
 \multirow{4}{*}{W4A8} & QoQ              & 77.64 & 72.81 & 43.60 & 74.00 & 68.03 & 67.22 \\
                        & QoQ-g128         & 78.07 & 73.32 & 44.80 & 74.98 & 68.59 & 67.95 \\
                        & QQQ              & 77.42 & 69.15 & 42.15 & 73.54 & 65.98 & 65.65 \\
                        & QQQ-g128         & 78.51 & 72.94 & 44.37 & 74.53 & 67.01 & 67.47 \\
\cmidrule(lr){1-8}
 W4Ax & Atom-g128        & 75.14 & 52.99 & 38.40 & 69.37 & 62.75 & 59.73 \\
\cmidrule(lr){1-8}
 \multirow{2}{*}{W4A4} & \textbf{APEX4}-g128      & 75.50 & 67.00 & 38.10 & 72.16 & 62.40 & 63.73 \\
                        & \textbf{APEX4}-mix       & 74.59 & 66.96 & 37.54 & 67.97 & 64.60 & 62.73 \\
\midrule
\multicolumn{8}{l}{\textbf{Llama-2-13B}} \\
\cmidrule(lr){1-8}
 FP16 & --                    & 80.52 & 77.44 & 49.06 & 79.38 & 72.22 & 71.72 \\
\cmidrule(lr){1-8}
 \multirow{4}{*}{W4A8} & QoQ              & 79.71 & 75.97 & 48.38 & 77.80 & 70.96 & 70.56 \\
                        & QoQ-g128         & 79.43 & 77.06 & 48.81 & 78.35 & 70.48 & 70.83 \\
                        & QQQ              & 79.43 & 74.75 & 48.12 & 77.27 & 70.32 & 69.98 \\
                        & QQQ-g128         & 79.98 & 76.64 & 48.55 & 78.63 & 71.82 & 71.13 \\
\cmidrule(lr){1-8}
 W4Ax & Atom-g128        & 76.50 & 57.49 & 42.32 & 73.84 & 67.40 & 63.51 \\
\cmidrule(lr){1-8}
 \multirow{2}{*}{W4A4} & \textbf{APEX4}-g128      & 78.61 & 76.13 & 44.62 & 75.32 & 64.70 & 67.87 \\
                        & \textbf{APEX4}-mix       & 78.34 & 75.58 & 44.19 & 74.44 & 69.53 & 68.41 \\
\bottomrule
\end{tabular}
}
\end{table}
\textbf{Perplexity.}
Table~\ref{tab:ppl_wikitext2} reports WikiText2 perplexity across ten models from four families.
APEX4-g128 incurs a perplexity increase of 0.38--1.65 over FP16.
Within the Llama-2 family, degradation shrinks consistently with scale
(+0.97, +0.71, +0.63 from 7B to 70B), while the larger increases on
Llama-3 and Qwen2.5 reflect their higher quantization difficulty.
Compared to Atom-g128, which employs W4Ax mixed-precision
(retaining outlier activations at INT8),
our pure W4A4 APEX4-g128 achieves comparable or lower perplexity
on six of seven evaluated Llama models (e.g., 6.09 vs.\ 6.12 on Llama-2-7B, 7.70 vs.\ 7.76 on Llama-3-8B),
despite operating entirely in INT4 without mixed-precision fallback.
Relative to W4A8 methods (QoQ-g128, QQQ-g128), APEX4-g128 shows a 0.20--1.06 perplexity gap,
consistent with the expected cost of reducing activation precision
from 8-bit to 4-bit. APEX4-mix trades slightly higher perplexity
for improved hardware efficiency on high-$\rho$ GPUs.
The perplexity increase over APEX4-g128 is 0.01--0.22
across Llama models and 0.12--0.30 across Qwen2.5 models,
a moderate cost that enables the per-channel routing
exploited in the end-to-end experiments of Section~\ref{sec:e2e}.

\textbf{Zero-shot accuracy.}

Table~\ref{tab:zeroshot_performance} reports accuracy across five tasks on Llama-2-7B and Llama-2-13B. APEX4-g128 incurs an average accuracy loss of 5.25\% on Llama-2-7B
and 3.85\% on Llama-2-13B relative to FP16.
Compared to Atom-g128, our method outperforms by 4.00\% (7B) and 4.4\% (13B),
indicating that activation smoothing and block-wise distillation
are more effective than Atom's dynamic outlier promotion
for pure 4-bit activation quantization. APEX4-mix performs comparably to APEX4-g128 on Llama-2-13B
(68.41\% vs.\ 67.87\%),
suggesting that per-channel quantization on less sensitive layers
does not necessarily degrade task accuracy.

\subsection{Kernel Performance Analysis}
\label{sec:kernel_eval}
\begin{figure*}[!htb]
    \centering
    \includegraphics[width=\textwidth]{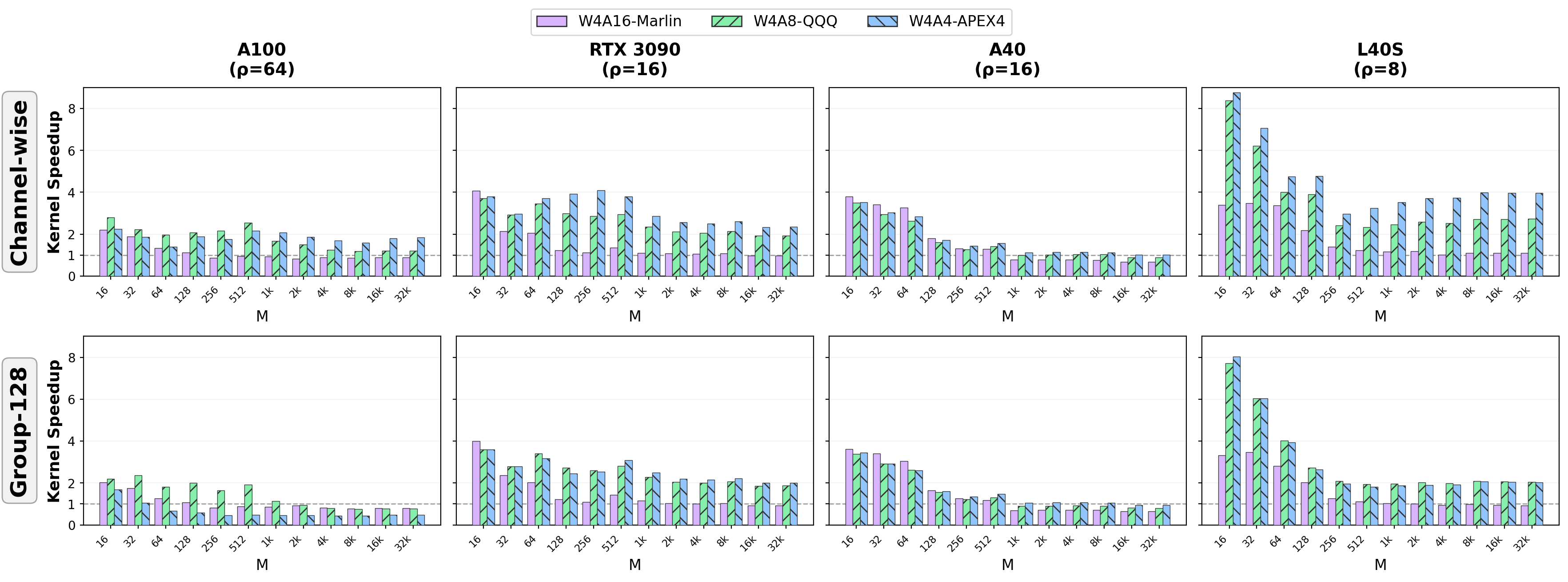} 
    \caption{Kernel speedup comparison across different precisions on four GPUs: A100, RTX 3090, A40, and L40S. W4A16 uses the Marlin~\cite{marlin} kernel, W4A8 uses the QQQ~\cite{QQQ} kernel, and W4A4 uses our kernel. All results are normalized to the FP16 baseline.}
    \label{fig:kernel_speedup}
\end{figure*}

The kernel level results under group-128 quantization reveal a clear $\rho$-dependent performance hierarchy, as shown in Fig.~\ref{fig:kernel_speedup}. 
On GPUs with $\rho \leq 16$, APEX4-g128 consistently outperforms FP16 cuBLAS baselines:
RTX~3090 ($\rho{=}16$) achieves $2.0$--$2.5\times$ speedup at large $M$,
while L40S ($\rho{=}8$) sustains $1.9$--$2.1\times$ in the compute-bound regime
and peaks at $8.0\times$ under memory-bound conditions ($M{=}16$).
In contrast, A100 ($\rho{=}64$) delivers only $0.43$--$0.47\times$ at large $M$,
as per-group dequantization saturates its scarce CUDA Cores.

Compared to APEX4-channel quantization, APEX4-g128 introduces uniform degradation across all GPUs,
but the magnitude scales with $\rho$:
on L40S the large-$M$ speedup drops from $3.7$--$4.0\times$ to $1.9$--$2.1\times$;
on A100 it collapses from $1.6$--$1.9\times$ to below $0.5\times$.

Comparing the three group-128 kernel variants, W4A16-g128 (Marlin), W4A8-g128 (QQQ), and W4A4-g128 (APEX4), reveals that the relative ranking depends on the compute regime.
At small $M$ where execution is memory-bound,
all three share the same 4-bit weight and thus similar memory-access cost;
W4A16-g128 achieves the highest speedup on RTX~3090 and A40
because its dequantization path is the simplest.
As $M$ grows and execution becomes compute-bound,
the ranking on $\rho \leq 16$ GPUs shifts to W4A4-g128 $>$ W4A8-g128 $>$ W4A16-g128,
with W4A4-g128 providing 10--20\% additional speedup over W4A8-g128.
On A100, however, this ranking inverts:
W4A8-g128 ($0.76$--$2.36\times$) consistently surpasses W4A4-g128 ($0.43$--$1.68\times$).
The divergence stems from the distinct dequantization placement:
W4A16-g128 and W4A8-g128 dequantize INT4 weights to higher precision before the MMA instruction, with group scaling fused into this step and largely hidden by MMA latency;
whereas W4A4-g128 executes MMA in INT4 and must apply per-group scale factors after every sub-tile iteration, serializing with MMA
and creating the CUDA Cores bottleneck.


\subsection{End-to-End Inference Performance}
\label{sec:e2e}
\begin{figure}[!htb]
    \centering
    \includegraphics[width=\textwidth]{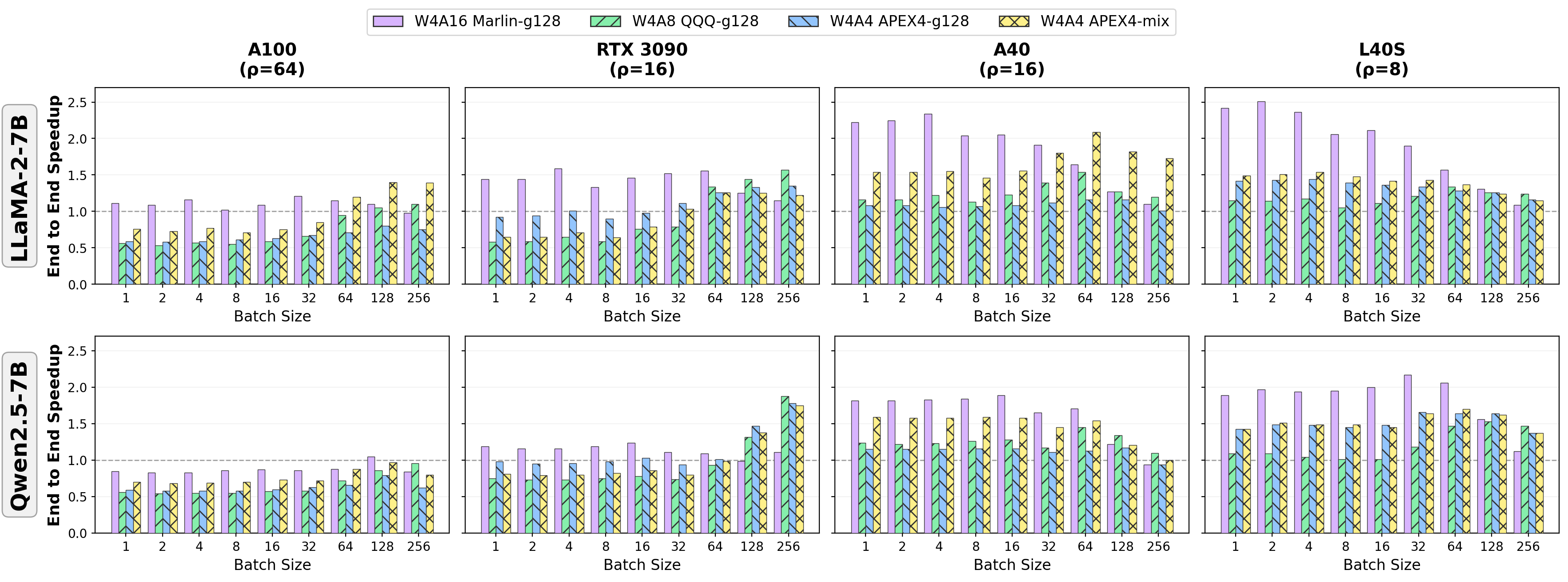} 
    \caption{Comparison of end-to-end speedup across different precisions on four GPUs: A100, RTX 3090, A40, and L40S. W4A16 uses the Marlin-g128~\cite{marlin} kernel, W4A8 uses the QQQ-g128~\cite{QQQ} kernel, and W4A4 uses our kernel. All results are normalized to the FP16 baseline.}
    \label{fig:endtoend}
\end{figure}

To examine whether the $\rho$-dependent hierarchy observed at the kernel level persists under realistic inference conditions,
we deploy APEX4 within unmodified vLLM and measure end-to-end throughput on Llama-2-7B and Qwen2.5-7B, isolating the effect of  our W4A4 kernels from other serving optimizations.
The absolute speedup numbers should therefore be interpreted
as a controlled isolation of the GEMM kernel's contribution
rather than production deployment benchmarks.
Our W4A4 kernel already employs a software pipeline
to overlap data loading with tensor core computing,
so the performance differences reported below
predominantly reflect the serial dequantization overhead.

The $\rho$-dependent ranking propagates to end-to-end inference. As shown in Fig.~\ref{fig:endtoend}, 
on $\rho \leq 16$ GPUs, APEX4-g128 delivers end-to-end speedup over FP16
across both models, broadly tracking the kernel-level pattern.
L40S ($\rho{=}8$) achieves the most consistent gains:
$1.16$--$1.44\times$ on Llama-2-7B and $1.37$--$1.66\times$ on Qwen2.5-7B.
RTX~3090 ($\rho{=}16$) exhibits a batch-size-dependent transition,
hovering near parity at small batch sizes but climbing to
$1.35\times$ (Llama-2-7B) and $1.78\times$ (Qwen2.5-7B) at BS$=$256.
A40 ($\rho{=}16$) shows a more constrained picture: APEX4-g128 remains above $1.0\times$ for Llama-2-7B ($1.01$--$1.16\times$)
but drops below parity on Qwen2.5-7B at large batch sizes
(e.g., $0.94\times$ at BS$=$256),
reflecting the narrower kernel-level margin on this GPU.
On A100 ($\rho{=}64$), APEX4-g128 remains below $1.0\times$ in all measured configurations:
$0.59$--$0.80\times$ on Llama-2-7B, $0.58$--$0.79\times$ on Qwen2.5-7B.
This is consistent with QServe's~\citep{Qserve} independent report
that existing W4A4 systems (Atom, Quarot) are 20--25\% slower
than TensorRT-LLM-W8A8 on A100,
although QServe did not attribute the root cause to the TC/CC throughput ratio.

APEX4-mix validates $\rho$-aware granularity assignment, routing most layers through per-channel kernels and yielding a markedly different profile.
On A100, APEX4-mix achieves $1.20$--$1.40\times$ on Llama-2-7B at BS$\geq$64,
compared to APEX4-g128's $0.71$--$0.80\times$ in the same regime.
On Qwen2.5-7B the A100 recovery is smaller ($0.80$--$0.97\times$ at BS$\geq$64),
which we attribute to differences in model architecture
(hidden dimensions, layer shapes) affecting per-layer kernel efficiency.
On A40, APEX4-mix achieves the highest speedup among all four methods,
reaching $2.09\times$ on Llama-2-7B (BS$=$64),
as A40's per-channel W4A4 kernels are already efficient at $\rho{=}16$.
On L40S, APEX4-mix and APEX4-g128 perform comparably
($1.15$--$1.54\times$ vs.\ $1.16$--$1.44\times$ on Llama-2-7B),
as both configurations benefit from the low $\rho$.

At small batch sizes, inference is memory-bound:
all schemes load the same 4-bit weights,
and INT4 Tensor Core throughput has limited opportunity to manifest.
W4A16-Marlin, with the lightest dequantization path,
tends to be the fastest variant at small batch sizes---reaching
$2.51\times$ on L40S and $2.34\times$ on A40 for Llama-2-7B.
As batch sizes increase and execution becomes increasingly computationally bottlenecked,
GPUs with p $\leq$ 16 alleviate the bottleneck of group dequantization and offer the throughput advantage of lower-bit Tensor Cores,
allowing the APEX4-g128, APEX4-mix, and W4A8-QQQ to achieve or even surpass the performance of the W4A16-Marlin.
However, on the A100, the dequantization overhead of the APEX4-g128 consistently diminishes its Tensor Cores advantage across all batch sizes tested.

For L40S, the APEX4-g128 kernel speedup of $1.9$--$2.1\times$
reduces to $1.16$--$1.66\times$ at the system level, indicating headroom for complementary techniques such as KV-cache quantization and fused attention~\citep{Qserve, COMET, ATOM}. These optimizations are largely orthogonal to our work: they address non-GEMM overhead, whereas $\rho$-aware granularity selection targets the CUDA-core bottleneck within the GEMM kernel.

\subsection{Dequantization Bottleneck Trends}
\begin{figure}[!htb]
    \centering
    \includegraphics[width=\columnwidth]{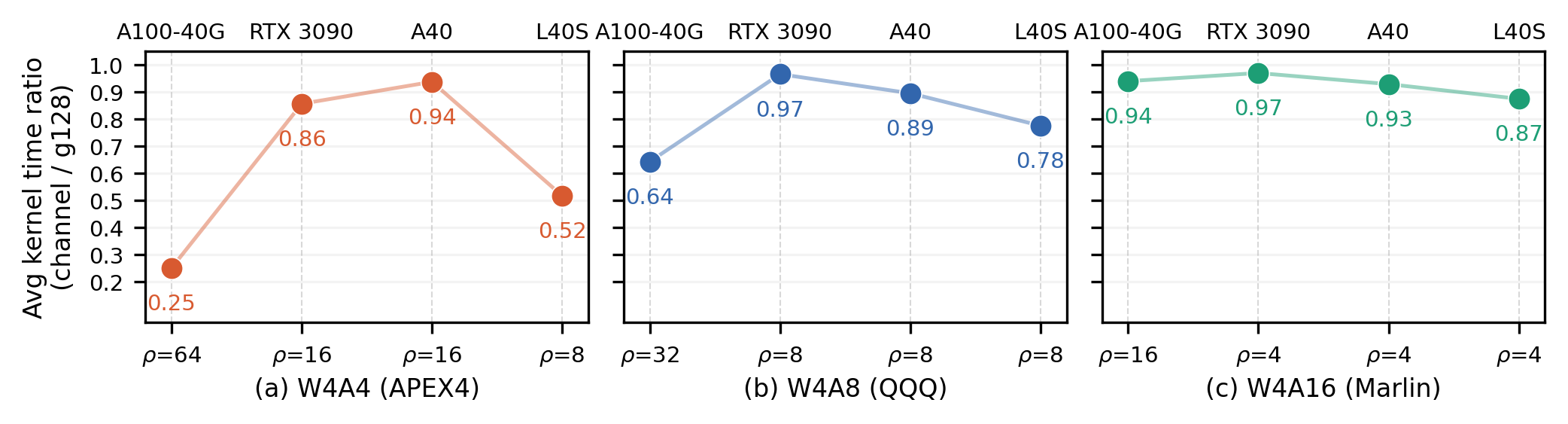} 
    \caption{Average kernel time ratio of channel to group-128. Each data point is the mean ratio across six GEMM sizes with $M \in \{1024, 2048, 4096, 8192, 16384, 32768\}$ and $N = K = 8192$. }
    \label{fig:ratio}
\end{figure}

Fig.~\ref{fig:ratio} shows the average kernel time ratio of channel-wise to group-128 quantization across four GPUs under W4A4, W4A8, and W4A16 precisions. A lower ratio generally indicates that the dequantization overhead introduced by group quantization is more pronounced.

The three subplots exhibit a largely consistent trend: as $\rho$ increases, the ratio tends to decrease. This trend is evident in W4A4, the ratio on the A100 ($\rho{=}64$) is approximately 0.25, whereas on the L40S ($\rho{=}8$) it approaches 0.52. In contrast, the W4A16 shows all GPU ratios clustered within the 0.87–0.97 range, with relatively small variation. This suggests that on GPUs with higher $\rho$, where Tensor Core throughput significantly exceeds that of CUDA Cores, the dequantization stage tends to become a more prominent bottleneck; meanwhile, lower-precision activations (INT4 $>$ INT8 $>$ FP16) appear to further amplify this effect.

It is worth noting, however, that GPUs sharing the same $\rho$ (e.g., RTX 3090 and A40, both with $\rho{=}16$) still exhibit different ratios, likely due to differences in memory bandwidth, L2 cache capacity, SM count, and other hardware characteristics. Therefore, this Fig.~\ref{fig:ratio} reflects each GPU's \emph{overall tolerance} for the dequantization overhead of group quantization, rather than a strict functional relationship between $\rho$ and bottleneck severity. Nevertheless, $\rho$ remains a useful indicator for characterizing the severity of the dequantization bottleneck.

\section{Related work}
\label{sec:related}

Weight-activation quantization reduces inference cost by quantizing both weights and activations to low-bit precision, enabling integer tensor core instructions. Existing methods target W4A8 or W4A4 precision, both contending with per-group dequantization overhead on CUDA Cores.
For example, 
QServe~\citep{Qserve} adopts per-channel quantization on A100 but per-group on L40S, explicitly attributing this to the latter's stronger CUDA Cores throughput---a direct demonstration that hardware resource asymmetry influences granularity selection.
QQQ~\citep{QQQ} implements efficient W4A8 group quantization by transforming it into a channel-level formulation to reduce overhead.
LiquidGEMM~\citep{liquidgemm} pinpoints the throughput mismatch between CUDA Cores dequantization and tensor cores, and proposes an implicit fine-grained pipeline overlapping weight loading, dequantization, and MMA across warp groups.
For W4A4 quantization,
Atom~\citep{ATOM} promotes activation outliers to 8-bit, requiring mixed-precision tensor core scheduling.
COMET~\citep{COMET} supports mixed W4A4/W4A8 via fine-grained SM scheduling but still dequantizes weights to 8-bit internally.
QuaRot~\citep{Quarot} eliminates activation outliers via fused Hadamard transforms, enabling fully end-to-end 4-bit quantization.

These works identify per-group dequantization as a key performance bottleneck, and recent work~\citep{Chen25} shows that finer granularity decreases quantization error but introduces overhead often offsetting low-bit gains. However, no prior work systematically analyzes how intra-SM resource heterogeneity impacts this bottleneck. We address this gap by establishing an analytical framework linking intra-SM hardware resource distribution to group-wise quantization efficiency, and validate it with a pure W4A4 kernel across multiple GPU architectures.

\section{Conclusion}
\label{sec:conclusion}
This paper identifies the intra-SM Tensor-Core-to-CUDA-Core throughput ratio $\rho$ as the primary hardware factor governing the practical efficiency of W4A4 group quantization. Through controlled benchmarks across four GPUs spanning two architecture generations, we show that the same W4A4-g128 kernel ranges from $2.5\times$ speedup on low-$\rho$ platforms to $0.43\times$ on high-$\rho$ ones, establishing that W4A4 viability is platform-dependent rather than universally infeasible. Guided by this analysis, we co-design an activation-smoothing and block-wise distillation algorithm with optimized pure INT4 GEMM kernels, and integrate them into APEX4, which adapts quantization granularity to the target GPU's $\rho$. End-to-end serving benchmarks confirm that $\rho$-aware configuration makes W4A4 a practical inference primitive across architecturally diverse GPU deployments.

\bibliography{tmlr}
\bibliographystyle{tmlr}

\end{document}